\documentclass[sn-mathphys,Numbered]{sn-jnl}

\usepackage{graphicx}%
\usepackage{multirow}%
\usepackage{amsmath,amssymb,amsfonts}%
\usepackage{amsthm}%
\usepackage{mathrsfs}%
\usepackage[title]{appendix}%
\usepackage{xcolor}%
\usepackage{textcomp}%
\usepackage{manyfoot}%
\usepackage{booktabs}%
\usepackage{algorithm}%
\usepackage{algorithmicx}%
\usepackage{algpseudocode}%
\usepackage{listings}%

\usepackage{amsmath}
\usepackage{MnSymbol}%
\usepackage{wasysym}%
\usepackage{tabularray}
\usepackage{arydshln}

\usepackage{color}
\usepackage{ulem}
\usepackage{lineno,hyperref}
\modulolinenumbers[5]
\usepackage{xcolor}
\usepackage{epstopdf, epsfig}
\usepackage[export]{adjustbox}
\usepackage{caption}
\usepackage{subcaption}
\usepackage[english]{babel}

\usepackage{mathMacros}

\theoremstyle{thmstyleone}%
\newtheorem{theorem}{Theorem}
\newtheorem{proposition}[theorem]{Proposition}%

\theoremstyle{thmstyletwo}%
\newtheorem{example}{Example}%
\newtheorem{remark}{Remark}%

\theoremstyle{thmstylethree}%
\newtheorem{definition}{Definition}%

\begin{document}

\title[Article Title]{General hydrodynamic features of elastoviscoplastic fluid flows through randomised porous media}


\author[1]{\fnm{Saeed} \sur{Parvar}}

\author*[2]{\fnm{Emad} \sur{Chaparian}}\email{emad.chaparian@strath.ac.uk}

\author[1]{\fnm{Outi} \sur{Tammisola}}

\affil[1]{SeRC and FLOW, Engineering mechanics, KTH Royal Institute of Technology, SE-10044 Stockholm, Sweden}

\affil[2]{James Weir Fluid Laboratory, Department of Mechanical \& Aerospace Engineering, University of Strathclyde, Glasgow, United Kingdom}


\abstract{A numerical study of yield-stress fluids flowing in porous media is presented. The porous media is randomly constructed by non-overlapping mono-dispersed circular obstacles. Two class of rheological models are investigated: elastoviscoplastic fluids (i.e.~Saramito model) and viscoplastic fluids (i.e.~Bingham model). A wide range of practical Weissenberg and Bingham numbers is studied at three different levels of porosities of the media. The emphasis is on revealing some physical transport mechanisms of yield-stress fluids in porous media when the elastic behaviour of this kind of fluids is incorporated. Thus, computations of elastoviscoplastic fluids are performed and are compared with the viscoplastic fluid flow properties. At a constant Weissenberg number, the pressure drop increases both with the Bingham number and the solid volume fraction of obstacles. However, the effect of elasticity is less trivial. At low Bingham numbers, the pressure drop of an elastoviscoplastic fluid increases compared to a viscoplastic fluid, while at high Bingham numbers we observe drag reduction by elasticity. At the yield limit (i.e.~infinitely large Bingham numbers), elasticity of the fluid systematically promotes yielding: elastic stresses help the fluid to overcome the yield stress resistance at smaller pressure gradients. We observe that elastic effects increase with both Weissenberg and Bingham numbers. In both cases, elastic effects finally make the elastoviscoplastic flow unsteady, which consequently can result in chaos and turbulence.}




\maketitle

\section{Introduction}\label{sec:intro}

Fluid flow through porous media is one of the classic fluid mechanics problems due to its importance for many industrial and environmental applications. Yield-stress fluids flowing in porous media is also intrinsic to a wide range of processes from cement grouting, enhanced oil recovery, hydraulic fracturing, etc.~ in construction and oil \& gas industries to vertebroplasty (i.e.~injection of bone ``cement" into fractured vertebra) \cite{williams2024bone}.

Over the past decades, major forward steps have been made in revealing the features of the flow by connecting the micro-scale features to the macro-scale hydrodynamic properties by integrating the linear equations governing the viscous fluids based on Darcy's findings. His framework was a baseline for further progressions in the fluid mechanics of porous media when the working fluid is Newtonian.

However, following experiments and theoretical works highlighted the invalidity of such approaches when it comes to more complex fluids such as viscoelastic and yield-stress fluids due to the non-linearity of the constitutive equations.
More specifically for the yield-stress fluids, which are the focus of the present study, the issue is two-fold:
\begin{itemize}
\item[(i)] Yield-stress fluids rheologically behave like solids when the applied stress is less than a threshold---the yield stress. If we translate this to porous media terminology, it means that if the applied pressure gradient is less than a threshold, then there is no flow inside the porous media. Hence, a {\it finite} pressure gradient is required to initiate the flow of a yield-stress fluid inside a porous medium.
\item[(ii)] Beyond the above limit (i.e.~the yield limit), due to non-linearity of the constitutive equations, Darcy's approach is not valid. Hence, permeability does not only depend on the geometrical features of the porous media (e.g.~porosity and shape of the obstacles/void spaces), but also it depends on the rheological parameters of the fluid. Consequently, we cannot decouple fluid's rheology and pore geometry which is feasible in Darcy's approach for Newtonian fluids.
\end{itemize}
These features have been demonstrated in several theoretical, numerical and experimental studies \cite{talon2013determination,bauer2019experimental,chaparian2019porous,liu2019darcy,chaparian2021sliding,chaparian2024percolation} and three main regimes have been identified for yield-stress fluid flows through porous media knowing that when the applied pressure gradient is less than the critical pressure gradient, there is no flow. If the applied pressure gradient slightly exceeds the critical value, the flow is extremely localised and goes through a single channel \cite{liu2019darcy,fraggedakis2021first}. In this regime, the flow rate linearly scales with the excessive pressure gradient. As the applied pressure gradient increases, a second regime emerges in which more and more channels will appear and the flow rate scales quadratically with the excessive applied pressure gradient. In the third regime, where the applied pressure gradient is much higher than the critical value, the flow rate again scales linearly with the excessive pressure gradient. Talon and co-workers in a series of studies \cite{talon2013determination,bauer2019experimental,talon2022determination} distinguished these three distinct regimes in the yield-stress fluid flow through porous media, which have been validated further by Chaparian \& Tammisola \cite{chaparian2021sliding} in randomised porous media.

Emerging sophisticated experimental techniques in the recent years have revealed that ``simple" yield-stress fluid rheological models are not sufficient for describing complex hydrodynamic features of many practical yield-stress fluids \cite{putz2008settling,holenberg2012particle,villalba2023}. Therefore, a number of rheological models have been proposed which include thixotropic and/or elastic behaviour of the fluid as well. A minimal elastoviscoplastic model is proposed by Saramito \cite{saramito2007new} which is indeed a combination of the Bingham and the Oldroyd-B models and is capable of capturing more crucial facets of complex yield-stress fluid flows \cite{fraggedakis2016soft,fraggedakis2016JNNFM,villalba2023}. Hence, we use this rheological model in the present study.

We previously analysed the elastoviscoplastic fluid flow at micro-scale inside model porous media, i.e.~flow over obstacles in a single periodic cell   \cite{de2018elastoviscoplastic,chaparian2019porous}. We found that for relatively small Bingham numbers, the slightly elastic behaviour of the fluid increases the pressure gradient. However, at moderate Bingham numbers, the pressure gradients for viscoplastic and weakly elastoviscoplastic fluids were found to be very similar. The elasticity, nevertheless, changes the shape of unyielded and fouling regions to some extent. Indeed, in symmetric model geometries, given that the inertia is negligible, the unyielded/fouling regions shape is symmetric for viscoplastic fluids, while it is asymmetric for elastoviscoplastic fluids. This fact has been thoroughly discussed in \cite{chaparian2019adaptive,villalba2023}. 

Here, the focus is again on elastoviscoplastic fluid flows in porous media. The objectives are: (i) pushing the analysis further to a wider range of the Bingham numbers (i.e.~yield limit) and porosities; (ii) addressing the effect of the Weissenberg number (i.e.~the extent of elastic behaviour of the yield-stress fluid) and more importantly, (iii) investigating the randomised porous media (rather than a periodic pore-scale cell) and also bridging our previous micro-scale study to macro-scale features. Moreover, we shed some light on how the elastic effects make the flow oscillatory by time which can eventually result in elastic instabilities.

Elastic turbulence found some growing attention during the past decades \cite{Steinberg2021,datta2022perspectives} which is attributed to the influence of polymer chains in the creeping flow regime where the inertial forces are negligible. This results in chaotic patterns which are absent in corresponding creeping flow of Newtonian fluids. In elastic turbulence, flow behavior is mainly governed by non-linear elastic stresses characterised by large Weissenberg numbers ($1 \ll Wi$) or consequently large elasticity numbers which has a wide range of applications, spanning from materials science to the biomedical engineering \cite{Steinberg2021,PARVAR20212}.

For several yield-stress fluid flows, it has been reported that high elastic effects can be observed even at low Weissenberg numbers if the Bingham number is large enough \cite{chaparian2019adaptive,Izbassarov2020,Izbassarov2021}. This fact resulted in proposing $Wi \times Bn$ as the effective parameter controlling the emergence of elastic effects by Chaparian \& Tammisola \cite{chaparian2019adaptive} which has been experimentally demonstrated as well by Villalba et al.~\cite{villalba2023}. Indeed, synergy of elasticity and plasticity of the fluid triggers elastic effects even at small Weissenberg numbers. Hence, we expect that alternative ``turbulence" happens in elastoviscoplastic fluid flows where both Reynolds and Weissenberg numbers are small while the Bingham number is fairly large. This has been slightly discussed here for flows in porous media although it is not the main aim of the present study.

In what follows, we set the problem in section \ref{section:Problem_setup} and briefly discuss the utilized numerical methods. The results are presented in section \ref{section:results} and conclusions \& discussions in section \ref{section:conclusion}.

\section{Problem setup}\label{section:Problem_setup}
In the present study, we investigate two-dimensional very low Reynolds number flows of yield-stress fluids (viscoplastic \& elastoviscoplastic) through porous media composed of mono-dispersed rigid circular obstacles ($X$) in a square domain ($\Omega$) of size $L \times L$. In this section, the governing equations, porous media construction and numerical methods are explained. To model the rheology of the elastoviscoplastic material, here we rely on Saramito model \cite{Saramito2007}, in which the material is described as a Kelvin-Voigt viscoelastic solid before yielding, and a Bingham fluid with an extra elastic memory after yielding.

\subsection{Mathematical formulation}
The non-dimensional continuity, momentum and constitutive equations are as follows:
\begin{equation}
  \displaystyle\boldsymbol{\nabla} \cdot \boldsymbol{\mathbf{u}} = 0,~~\text{in} ~\Omega/\bar{X},
\label{eq:cont}
\end{equation}

\begin{equation}
Re \left[ \pddt{\tens{u}} + ( \tens{u} \cdot \boldsymbol{\nabla}) \tens{u} \right] = 
-\boldsymbol{\nabla} p + 
\boldsymbol{\nabla} \cdot ( \btau + 
\beta \dot{\boldsymbol{\gamma}}), ~~\text{in} ~\Omega/\bar{X},
\label{eq:Cauchy}
\end{equation}

\begin{equation}
Wi~ \ucd{\btau} +
\left( 1-\frac{Bn}{\Vert \btau \Vert_{\text{v}}} \right)_+ \btau = (1-\beta) \dot{\boldsymbol{\gamma}},  ~~\text{in} ~\Omega/\bar{X},
\label{eq:const}
\end{equation}
where $t$ is time, ${\mathbf{u}}=(u,v)$ is the velocity vector, $p$ denotes the pressure, $\btau$ the stress tensor (sometimes is termed as {\it extra} stress tensor), and $\dot{\boldsymbol{\gamma}} =(\nabla\boldsymbol{\mathbf{u}}+\nabla \mathbf{u}^T)$ is the rate of deformation tensor where $\nabla\boldsymbol{\mathbf{u}}=\partial u_i/\partial x_j$. Here, $\Vert \cdot \Vert_{\text{v}}$ is the equivalent von Mises stress, $\left( \cdot \right)_+$ is the positive part of the argument (i.e.~is equal to zero if the argument is negative and otherwise is equal to the argument), and $ \overset{\triangledown}{\cdot} $ is the upper-convected derivative:
\begin{equation}
\ucd{\btau} \equiv \pddt{\btau} + (\tens{u} \cdot \boldsymbol{\nabla}) \btau - (\grad{\tens{u}})\, \btau - \btau\, ( \trans{\grad{\tens{u}}} ).
\label{eq:ucd}
\end{equation}

In these equations, the velocity vector is scaled with $\hat{U}$ (i.e.~the mean inlet velocity), pressure and stress tensor is scaled with the characteristic viscous stress $\hat{\mu} \hat{U}/\hat{D}$ where $\hat{D}$ (i.e.~diameter of the obstacles) is the length scale and $\hat{\mu}$ is the total viscosity. Moreover, $\beta = \hat{\mu}_s/\hat{\mu}$ where $\hat{\mu}_s$ is the solvent viscosity.

The non-dimensional parameters are the Reynolds number, Weissenberg number, and the Bingham number:
\[
Re = \frac{\hat{\rho} ~\hat{U} \hat{D}}{\hat{\mu}},~~Wi = \frac{\hat{\lambda} ~\hat{U}}{\hat{D}},~~Bn = \frac{\hat{\tau}_{y} \hat{D}}{\hat{\mu} ~\hat{U}}.
\]
The Reynolds number throughout this study is fixed and is equal to unity. The Weissenberg number represents as the ratio between elastic and viscous forces, and the Bingham number describes the ratio of the yield stress of the material to the characteristic viscous stress. {\color{red}In Appendix A, we discuss other possible scalings.}

Various approaches have been used in the numerical simulation of viscoelastic polymer solutions to handle the numerical instabilities appearing at ``high" Weissenberg numbers \cite{PARVAR20201,PARVAR20202,PARVAR20211,PARVAR20212,PARVAR20222}. In this study, the log-conformation approach, proposed by Fattal and Kupferman \cite{Fattal2004,Fattal2005}, is used. Although $Wi$ range studied here is not ``high", in elastoviscoplastic fluids the elastic features can appear even at relatively low Weissenberg numbers, see \cite{chaparian2019adaptive,Chaparian2020migration,villalba2023}.
For the case of the Saramito model (i.e.~equation \ref{eq:const}), the conformation tensor $\boldsymbol{A}$ reads,
\begin{equation}
\btau =\frac{1-\beta}{Wi}(\boldsymbol{A}-\boldsymbol{I}).
\label{eq:Kramers}
\end{equation}
For more details on log-conformation formulation, the readers are refereed to \cite{izbassarov2018computational}.
\subsection{Porous medium construction}\label{section:Porous_medium_construction}
As mentioned, here we are interested in 2D flows through porous media constructed of mono-dispersed rigid circular obstacles ($X$) in a domain ($\Omega$) of size $L \times L$. These obstacles have a unit diameter (i.e.~the length scale used in the above equations is the diameter of the obstacles) and $L=24$. The centres of obstacles are chosen entirely random with uniform distribution. The constraints are non-overlapping and also being completely in the computational box (i.e. they do not touch the borders of the computational domain).

\begin{figure} 
\centering
\includegraphics[width=\linewidth]{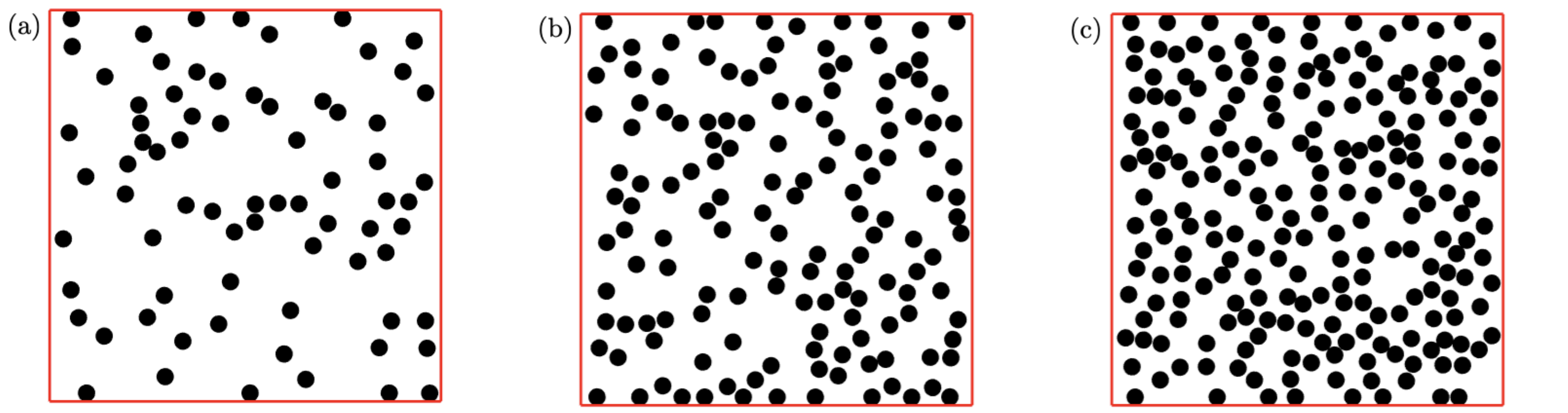}
\caption{Porous media: (a) $\phi=10\%$, (b) $\phi=20\%$, (c) $\phi=30\%$.}
\label{fig:schematic}
\end{figure}

A sketch illustrating the considered geometries is presented in Fig.~\ref{fig:schematic}. The ``volume fraction" of the obstacles is denoted by $\phi=\text{meas}(X)/\text{meas}(\Omega)=N\pi/4 L^2$ where $N$ is the number of obstacles. Hence, the porosity of the medium $\Phi$ (i.e.~void fraction) can be defined simply as $\Phi=1-\phi$. In the present study, we study three cases with the solid volume fraction $10\%, 20\%$, and $30\%$ (i.e.~porosity $\Phi=90, 80$ and $70\%$), as displayed in Fig.~\ref{fig:schematic}.

Although in the present study, due to high computational cost (especially the elastoviscoplastic problem), we do not follow a statistical approach to simulate the flow in many realisations, the position of the obstacles are chosen completely random in the studied cases to avoid any bias in the results. Indeed, all the results are reported based on the realisations shown in Fig.~\ref{fig:schematic} in which the obstacles are randomly generated to mimic the desired porosity.

\subsection{Highlights of the computational details}\label{section:Highlights_Computational}

In the following simulations, the no-slip boundary condition is imposed at the surface of the obstacles (${\mathbf{u}}=0$ on $\partial X$), and free-slip boundary conditions at the lateral (i.e.~top and bottom) boundaries, and the periodic boundary condition is implemented in the streamwise direction, while the left boundary is the inlet (i.e.~the fluid flows from left to right).

In what follows, we briefly explain the two computational methods used in the present study to solve elastociscoplastic \& viscoplastic problems. For more details, the readers are referred to our previous studies where we developed/implemented these methods. These references are mentioned accordingly.
\subsubsection{Viscoplastic fluid flow simulations}
We implement augmented Lagrangian method to simulate viscoplastic fluid flows \cite{roquet2003adaptive,roquet2008adaptive}. In this method, regularisation is avoided as the method is capable of handling the discontinuity in the Bingham model by relaxing the rate of the strain tensor. An open source finite element environment FreeFEM++ \cite{MR3043640} is used for discretization and meshing which has been widely validated in our previous studies \cite{chaparian2017yield,iglesias2020computing,chaparian2021sliding,chaparian2022vane,medina2023rheo}. Anisotropic adaptive mesh in $\Omega/\bar{X}$ is combined with this method to get smoother yield surfaces and ensure high resolution of the flow features \cite{chaparian2019adaptive,roquet2003adaptive}.
\subsubsection{Elastoviscoplastic fluid flow simulations}\label{sec:EVP_comp}

The evolution of the conformation tensor (i.e. Eqs.~\ref{eq:const} \& \ref{eq:ucd}) and the momentum equation are discritized using second-order central finite difference scheme on a uniform staggered grid in $\Omega$, except for the advection term in Eq.~\ref{eq:ucd} for which we use a fifth-order weighted essentially non-oscillatory (WENO) scheme \cite{Shu2009,Sugiyama2011}. The time integration is performed with a fractional-step, third-order explicit Runge–Kutta scheme \cite{Kim1985} for all equations. For more detail, the reader is referred to \cite{Izbassarov2018,Izbassarov2021}.

In the present study, the immersed boundary method (IBM) is used to represent the rigidity of obstacles (i.e.~$X$), coupled with the uniform computational grid $\Omega$, and hence exploit the possibility of utilizing efficient computational algorithms such as highly-scalable, fast Fourier transform (FFT)-based pressure solver. Hence in this way, the simulation is performed on a uniform Cartesian Eulerian grid with the solid surface reproduced by a uniformly distributed Lagrangian grid \cite{Breugem2012,Luo2007,Izbassarov2018,PARVAR20231}. The Eulerian grid is uniform in the two directions of the computational domain and the Lagrangian points are equally distributed on the solid surfaces---32 points per diameter equivalent to 101 points on the surface of each obstacle. The in-house developed IBM code has been widely verified in our previous physical studies involving multiphase elastoviscoplastic fluid flows such as flows in porous media \cite{Chaparian20202,de2018elastoviscoplastic}, particle migration \cite{Chaparian2020migration}, and flow past an obstacle \cite{Izbassarov2018,Sarabian2020,PARVAR20231}. 

\begin{table}[!h]
\def~{\hphantom{0}}
  \begin{tabular}{ccccc}
    $Wi$    & $Re$ & ~Hulsen et al.~\cite{hulsen2005flow}~ & ~Minaeian et al.~\cite{minaeian2022effects}~  &  ~Present simulations  \\[3pt]
    \hline
    \hline
       0.2 & 1 & 130.363 & --- & 133.31 \\
       2  & 2 & 118.471 & --- & 124.819 \\
       2  & 200 & ---  & 1.2006 & 1.21933 \\
  \hline
  \end{tabular}
  \caption{Drag coefficient of a cylinder in an infinite unidirectional viscoelastic fluid flow. Please note that $Wi$ and $Re$ in this table are defined based on the particle diameter, whereas the previous studies defined it based on the radius.}
  \label{tab:Cd}
\end{table}

For the present flow configuration, we provide an estimate of the drag coefficient around a single cylinder in viscoelastic flow obtained with the present IBM method, for time-independent and time-dependent flow cases. More comprehensive results are available in \cite{PARVAR20231}. It should be mentioned that some previous studies indicate that the accuracy of the IBM method in viscoelastic flow can sometimes be improved by the so-called smooth continuation which allows a higher accuracy of stress gradients at the boundary\cite{Stein2017,Stein2019}, and at $Wi=0.1$ such method could provide an error of $<0.1 \%$. However, the accuracy of the smooth IBM approach seemed to deteriorate with increasing Weissenberg number, and already at $Wi=1.4$ ($Wi=0.7$ by their definition) the error is $4.4\%$ with 128 points per cylinder surface (see Fig.~7 in \cite{Stein2019}), which is comparable to our IBM method error at $Wi=2$. In this study, we also refrain from the use of artificial diffusion, which may have significantly affected elastic instabilities \cite{Gupta2019} and would have been necessary if otherwise accurate spectral discretisations were employed as in \cite{Stein2019}. 

\section{Results}\label{section:results}
The emphasis in the present study is on studying the effect of the Weissenberg and Bingham numbers, and also the porosity of the medium on the hydrodynamic features. Hence, we fix $\beta =0.5$ in all the elastoviscoplastic simulations and to minimize the inertial effects, we specify $Re \approx 1$. Nevertheless, we vary the Bingham number in the range of 0 to 500. Knowing that the relaxation times of most practical yield-stress fluids are relatively small compared to the viscoelastic fluids \cite{hormozi2011stable,villalba2023}, we simulate the flows at relatively small Weissenberg numbers, $Wi=0.05 ~\&~ 0.5$, except at $\phi=10 \%$ where the extreme case $Wi=10$ is also studied.

\begin{figure}[!h] 
\centering
\includegraphics[width=1.0\linewidth]{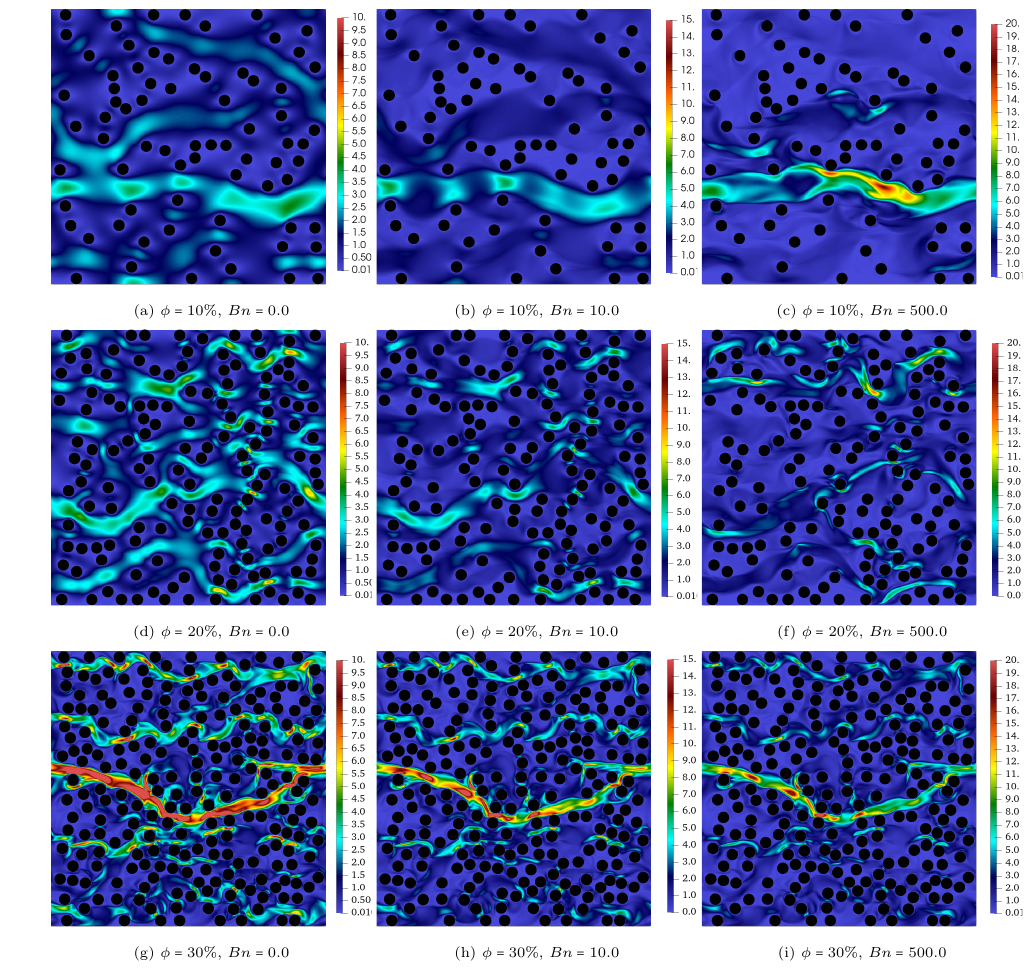}
\caption{\label{fig:Vel_PHI_10t30} The contour of velocity at $Wi=0.5$, $Bn=0,10,500$, and $\phi=10\%,20\%,30\%$.}
\end{figure}

\begin{figure}[!h] 
\centering
\includegraphics[width=1.0\linewidth]{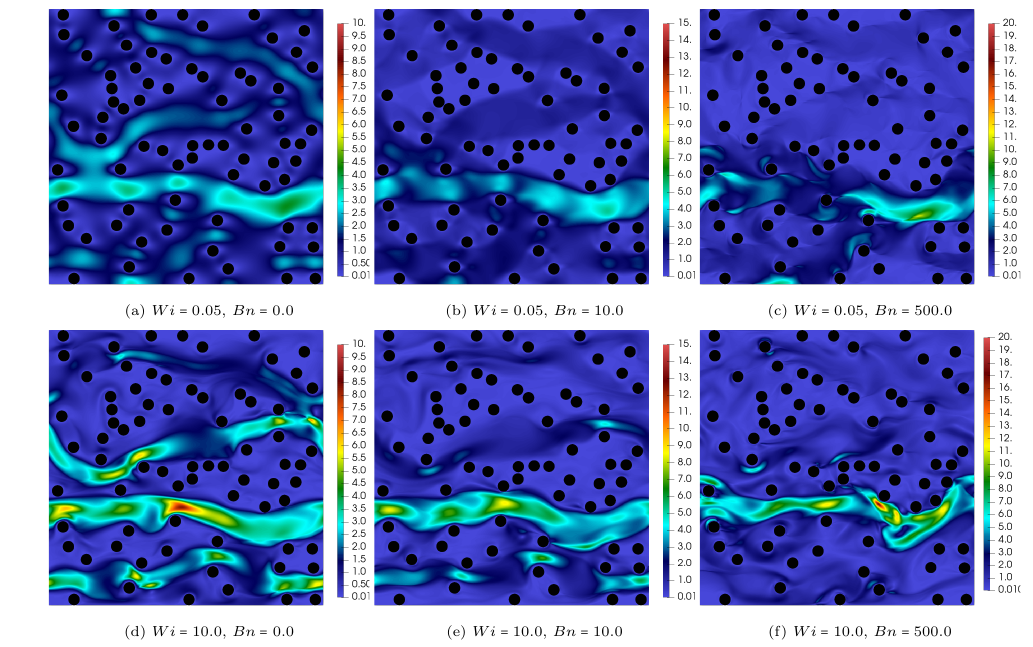}
\caption{\label{fig:Vel_Wi_005t10} The contour of velocity at $Wi=0.05~\&~10$ and $Bn=0,10,500$. The solid volume fraction is $\phi=10\%$.}
\end{figure}

The velocity contours (i.e.~$\vert \mathbf{u} \vert$) for the $\phi=10\%,20\%,30\%$ cases at $Wi=0.5$ and $Bn=[0,10,500]$ are presented in Fig.~\ref{fig:Vel_PHI_10t30}. Each row represents a constant solid volume fraction (i.e.~$\phi$) and it increases from top to bottom. Also, the Bingham number increases from the left to the right columns while the Weissenberg number is fixed at 0.5. Fig.~\ref{fig:Vel_Wi_005t10} shows again the velocity contour but the emphasis is on the effect of the Weissenberg number on the flow, hence at $\phi=10\%$, a wide range of $Wi$ is covered for different Bingham numbers.

An increase in the Bingham number from $Bn=0$ to $Bn=500$ results in a more confined flow or ``channelisation" which is well-documented before in the yield-stress fluids literature \cite{talon2013determination,liu2019darcy,chaparian2021sliding,fraggedakis2021first}. Indeed, at high Bingham numbers, a large portion of the fluid remains unyielded and the flow emerges in thin channels through the porous media due to the high yield stress of the fluid compared to the deriving stress imposed by the applied pressure gradient.

Although changing the Weissenberg number at a fixed Bingham number does not have significant effect on the contour of velocity (see Fig.~\ref{fig:Vel_Wi_005t10}), this increase results in appearing high velocity regions in the open channels when the fluid passes through the gaps between the obstacles which is intuitive for more elastic fluids. This becomes clear by comparing the top and bottom rows where the yellow and the red contours appear in the middle of the open channels for high $Wi$ flows (please note that the same colour ranges are used for the top and bottom counterparts).

\begin{figure}[!h] 
\centering
\includegraphics[width=1.0\linewidth]{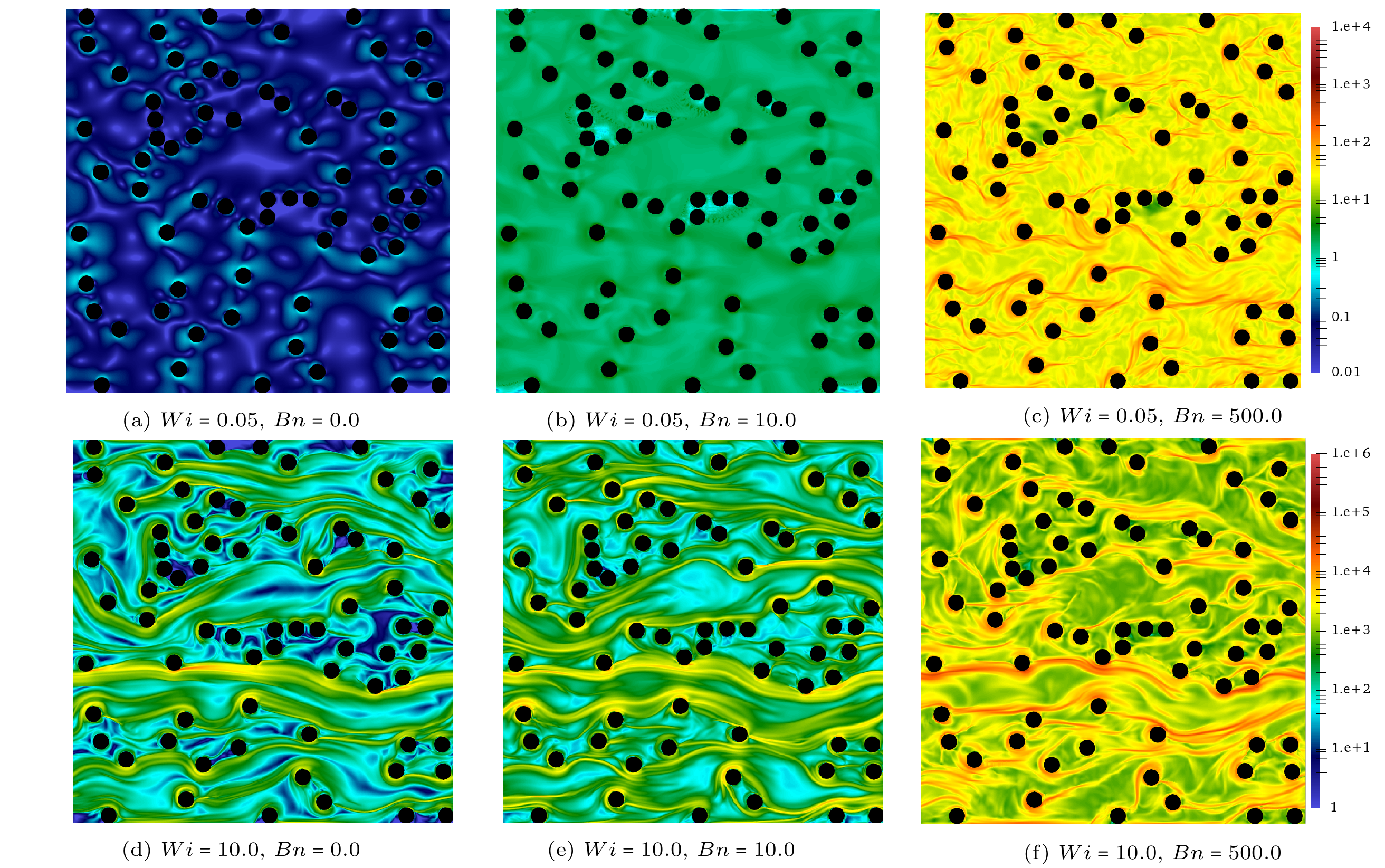}
\caption{\label{fig:Tra_PHI_10t30} The contour of the trace of the conformation tensor (i.e.~$A_{xx}+A_{yy}$) at $Wi=0.05~\&~10$ and $Bn=0,10,500$. The solid volume fraction is $\phi=10\%$.}
\end{figure}

\begin{figure}[!h] 
\centering
\includegraphics[width=1.0\linewidth]{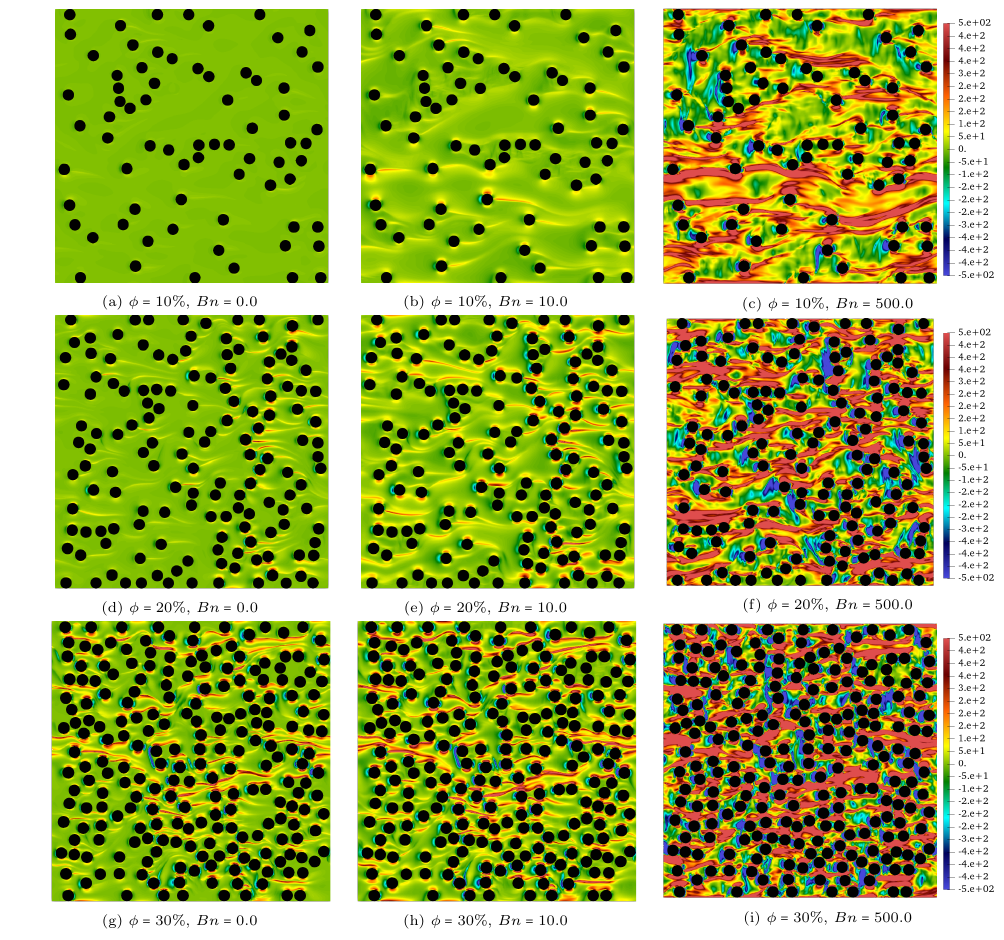}
\caption{\label{fig:N1_PHI_10t30} The contour of the normal stress difference (i.e.~$\tau_{xx}-\tau_{yy}$) at $Wi=0.5$, $Bn=0,~10~\&~500$, and $\phi= 10\%,~20\%~\&~30\%$.}
\end{figure}

To further analyse the effects of the Weissenberg and Bingham numbers on the flow features, Figs.~\ref{fig:Tra_PHI_10t30} and \ref{fig:N1_PHI_10t30} are plotted which display the contours of the trace of the conformation tensor (i.e.~$A_{xx}+A_{yy}$) and the normal stress difference (i.e.~$\tau_{xx}-\tau_{yy}$), respectively.

The trace of the conformation tensor(Fig.~\ref{fig:Tra_PHI_10t30}) is the sign of the elongation of the polymer chains in viscoelastic fluids. Knowing that the trace is zero for Newtonian and viscoplastic fluids (i.e.~the stress tensor is deviatoric), for elastoviscoplastic fluids, the trace of the stress tensor has the same interpretation as the viscoelastic fluids: the extent of stress generated by the elongation of fluid elements due to its elastic behaviour. As seen in Fig.~\ref{fig:Tra_PHI_10t30}, increasing the Weissenberg number from 0.05 (top panels) to 10 (bottom panels) results in an increase in the trace and also appearance of long and thin streaks in the streamwise direction. Indeed, longer streaks of high trace regions extend further downstream from the cylinders' surface. Moreover, increasing the Bingham number from $Bn=0$ to $Bn=500$ while the Weissenberg number is kept constant (compare top or bottom panels from left to right) leads to a more pronounced extension of fluid particles as indicated by the increase in the maximum of the trace, both locally and in the entire domain. For instance, when $Wi=0.05$ (top panels of Fig.~\ref{fig:Tra_PHI_10t30}), increasing the Bingham number from 0 to 500 changes the trace by almost three orders of magnitude: from dark blue (left panel) to yellow (right panel) contours. Notably, keeping the Weissenberg number the same and increasing the Bingham number leads to more elastic responses which is previously reported in \cite{chaparian2019adaptive,villalba2023} for complex internal flows of elastoviscoplastic fluids. It should be noted that also increasing the porosity leads to stronger streaks which is not shown here for the sake of brevity.

As mentioned, Fig.~\ref{fig:N1_PHI_10t30} depicts the contours of $\tau_{xx} - \tau_{yy}$. In the current study, it can be interpreted as a sign of localisation since indeed $\tau_{xx}$ is much higher than $\tau_{yy}$ when the normal stress difference is large. As it is clear from Fig.~\ref{fig:N1_PHI_10t30}, when $Bn=0$ (the Newtonian case), the entire fluid flows and $\tau_{xx} - \tau_{yy}$ is approximately negligible. However, at the other extreme, when $Bn=500$ (i.e.~moving towards the yield limit) and the flow is highly localised in one or few open channels, $\tau_{xx} - \tau_{yy}$ is quite large. It is also shown that the magnitude of $\tau_{xx} - \tau_{yy}$ increases with the increasing solid volume fraction. The value of $\tau_{xx} - \tau_{yy}$ is positive throughout the domain, except in the vicinity of the cylinders' upstream stagnation points. This can be interpreted as the transition of the flow direction from streamwise to vertical direction in order to align with the surface of the solid cylinders. Larger Bingham numbers and solid volume fractions increase the stagnation points effects and leads to an increase in the area with $\tau_{xx} - \tau_{yy} \leqslant 0$ shown by the blue colour contours.

\begin{figure}[!h]
\centering
\includegraphics[width=.7\linewidth]{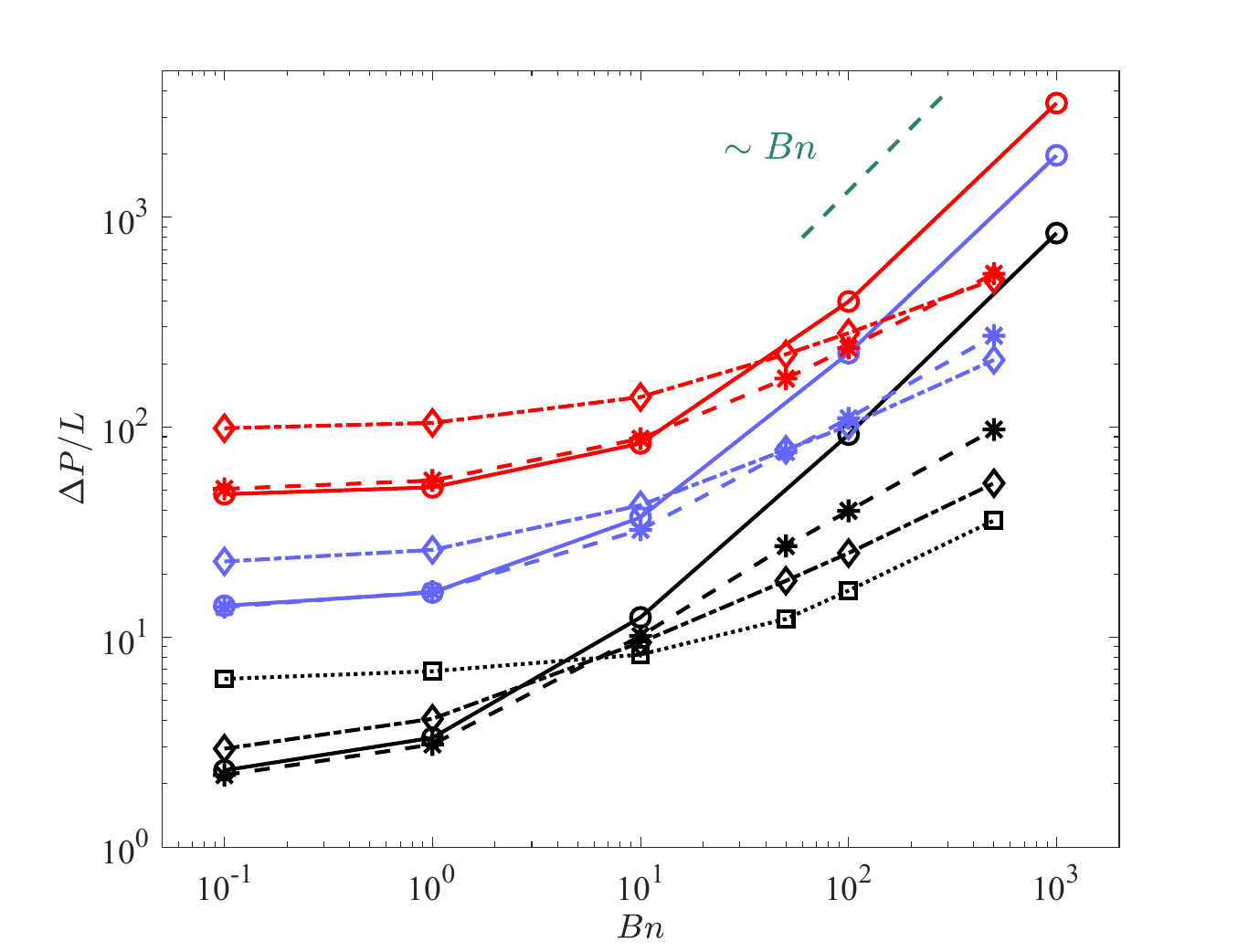}
\caption{Pressure gradient versus the Bingham number: solid lines \& circle symbols represent the viscoplastic simulations; dashed lines \& asterisks the cases of $Wi=0.05$; dashed dotted lines \& diamonds the cases of $Wi=0.5$; dotted line \& squares the case of $Wi=10$. The black colour stands for $\phi=10 \%$, the purple for $\phi=20 \%$, and the red for $\phi=30 \%$. The linear scaling is marked by green dashed line for reference.}
\label{fig:pressuredrop}
\end{figure}

To help us analyse the bulk transport properties, Fig.~\ref{fig:pressuredrop} shows the pressure gradient as a function of the Bingham number for different porosities (with different colours) and Weissenberg numbers (with different symbols \& line styles; see the caption). First of all, it is clear that for higher volume fractions the pressure drop is higher at the same $Wi$ and $Bn$ numbers, which is intuitive since the fluid should be percolated in a more packed arrangement of obstacles. At small Bingham numbers, the least studied Weissenberg number flow ($Wi=0.05$) and the viscoplastic flow are indistinguishable. However, as we increase the Bingham number and keep the Weissenberg number fixed at 0.05, we see that the discrepancy increases which has been evidenced in a number of our previous studies: plasticity of the fluid triggers the elastic behaviour of the material or indeed as $Wi \times Bn$ increases, we observe more elastic features in the flow \cite{chaparian2019adaptive,villalba2023}.

At lower $Bn$ values, if we increase the Weissenberg number, the pressure gradient increases; this is in agreement with our previous study {in a small periodic cell}, i.e.~\cite{chaparian2019porous}. At low Bingham numbers, the flow is rather evenly distributed over the porous medium, i.e.~the flow is not heterogeneous/localised and there are less unyielded/fouling regions. Hence, the drag increase as a function of the elasticity of the fluid at low Bingham numbers can be attributed to the non-straight streamlines where more energy is required to percolate an elastic fluid compared to a non-elastic one (due to stored elastic energy in fluid elements; see \cite{de2017viscoelasticb}). On the other hand, if we now focus on the high Bingham numbers regime, the effect of the elasticity of the fluid is inverse: indeed at high Bingham number regime, higher Weissenberg flows have less pressure gradient compared to the small Weissenberg flows. This is the regime which was not explored in our previous study \cite{chaparian2019porous}. The reason for drag reduction with elasticity at high Bingham numbers can be attributed to the fact that at higher Bingham numbers, the flow becomes more and more channelised (i.e.~localised in one or few open channels) and hence the streamlines are more straight. Thus, a qualitative argument can be made based on a flow through a channel, even though the analogy is not perfect. In a laminar channel flow, the Fanning friction factor decreases with the Weissenberg number \cite{Izbassarov2021} (see \cite{Chaparian2020migration} for the exact solution of elastoviscoplastic fluid Poiseuille flow). The main source of high drag in channel flow of yield-stress fluids is the sharp shear rate close to the walls which is mitigated by the elasticity of the fluid as the core plug region narrows down compared to the viscoplastic fluids. Thus, the elasticity of the fluid reduces the viscous stresses near the walls which results in less drag. More or less, the same mechanism can be observed here.

In summary, we observed two different regimes for drag. There is a zone of Bingham number in which a transition from {\it drag enhancement} to {\it drag reduction} happens, for example, for the case $\phi=10\%$, all curves associated with different $Wi$ meet around $Bn \approx 5$. Obviously, this zone shifts towards higher Bingham numbers as the solid volume fraction is increasing: for $\phi=20\%$, it is $Bn \approx 50$ and for $\phi=30\%$ it is $Bn \approx 500$. Nevertheless, it is clear that $Y_c$ is larger for higher Weissenberg flows. In other words, for the viscoplastic fluid (i.e.~case $Wi=0$), $\Delta P/L$ scales with the Bingham number (see \cite{chaparian2019porous} for the proof) while the slope is much more gentle at higher Weissenberg numbers as $Bn \to \infty$ (i.e.~at the yield limit). This has been generally observed before in other physical problems such as particle sedimentation in elastoviscoplastic fluids \cite{fraggedakis2016soft}: elastic stresses aid the particle to overcome the yield stress and sediment.

\begin{figure}[h] 
\centering
\includegraphics[width=1.0\linewidth]{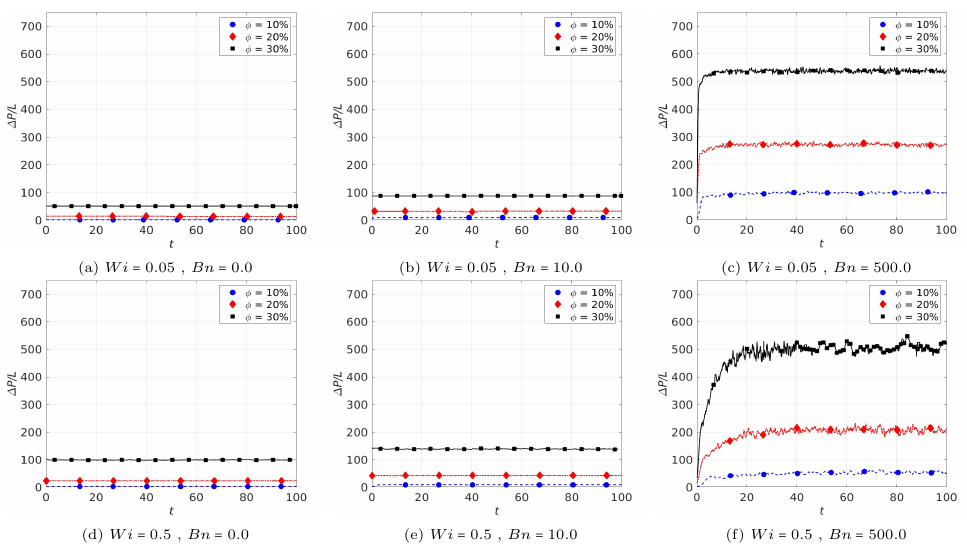}
\caption{\label{fig:Pre_Ins} Instantaneous pressure gradient ($\Delta P/L$ versus time) for $Wi=0.05~\&~0.5$ and $Bn=0,~10,~\&~500$.}
\end{figure}

\begin{figure}[h]
\centering
\includegraphics[width=.6\linewidth]{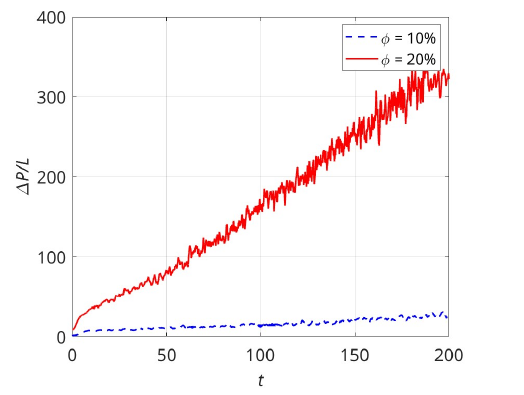}
\captionsetup{justification=centering}
\caption{\label{fig:Pre_Ins_NotC} Instantaneous pressure gradient ($\Delta P/L$ versus time) for the extreme case $Wi=10$ \& $Bn=500$ for different solid volume fractions.}
\end{figure}

It should be noted that the data represented in Fig.~\ref{fig:pressuredrop} associated with elastoviscoplastic fluid is the time-averaged pressure gradient. The instantaneous pressure gradient ($\Delta P/L$ versus time $t$) for different cases are plotted in Figs.~\ref{fig:Pre_Ins} and \ref{fig:Pre_Ins_NotC}. Clearly, when the Weissenberg and the Bingham numbers are relatively small (left and middle panels in Fig.~\ref{fig:Pre_Ins}), the steady state solutions are achieved after rather short time intervals irrespective of the solid volume fraction. The right panels (c,f) in Fig.~\ref{fig:Pre_Ins}, however, show the cases of high Bingham number $Bn=500$ (i.e.~large $Wi \times Bn$). In these cases, the pressure gradient oscillates around an average value even after long times. Hence, in these cases, only a {\it quasi-steady} state is reached by the elastoviscoplastic fluid. The average values in the final quasi-steady state are previously reported in Fig.~\ref{fig:pressuredrop}. As can be seen by comparing the different curves in Fig.~\ref{fig:Pre_Ins} (panel (f) in particular), the oscillations' amplitude is larger when the solid volume fraction is higher. Nevertheless, the oscillation amplitudes compared to the average values of $\Delta P/L$ are rather negligible. Another point is that, in these cases, the flow converges to the quasi-steady state solutions faster when the solid volume fraction is smaller: for the case $Wi=0.5~\&~Bn=500$ (see Fig.~\ref{fig:Pre_Ins}(f)), the flow reaches the quasi-steady state at $t \approx 5$ when $\phi=10\%$, at $t=10$ when $\phi=20\%$ and at $t=20$ when $\phi=30\%$.

If we increase the value of $Wi \times Bn$ even more (for instance $Wi=10$ \& $Bn=500$ which is represented in Fig.~\ref{fig:Pre_Ins_NotC}), there are cases in which the steady state solutions are not achievable, at least at practical time scales. This is especially pronounced at higher solid volume fractions. For example, the case $Wi=10, ~Bn=500~\&~\phi=0.2$ which is marked in Fig.~\ref{fig:Pre_Ins_NotC} by the red diamonds has not reached a plateau even at $t=200$ and the slope also does not seem to decrease over the next time decades. This is the main reason that $10 \leqslant Wi$ simulations are not reported in the present study when $0.1 < \phi$.

\section{Conclusions and discussion}\label{section:conclusion}

Numerical simulations of yield-stress fluid flows in porous media were performed. The focus was on comparing elastoviscoplastic (i.e.~Saramito model \cite{saramito2007new}) and viscoplastic (i.e.~Bingham fluid) rheological models. The porous media were constructed by randomised mono-dispersed non-overlapping circular obstacles where the size of the square computational domain is 48 times of the obstacle radius. Three different solid ``volume" fractions ($\phi$) are evaluated; $\phi=0.1$, 0.2 and 0.3. We presented a detailed analysis of the pressure drop for all of the case studies as a function of the Weissenberg number ($Wi$), Bingham number ($Bn$) and the porosity of the media ($1-\phi$).

It has been demonstrated that $\Delta P/L$ increases with the Bingham number regardless of the Weissenberg and $\phi$ which is completely intuitive since the yield stress resistance increases. Also, the pressure drop increases with the solid volume fraction of the medium at fixed any $Wi$ and $Bn$. However, pressure drop has a non-monotonic behaviour with the Weissenberg number: while at the small Bingham numbers, $\Delta P/L$ increases with the Weissenberg number, at large $Bn$, the Weissenberg number has an opposite effect. In other words, at low Bingham numbers, elasticity increases the pressure drop, while at high Bingham numbers, elasticity has a drag reduction effect. This is systematically observed for all the considered solid volume fractions.

At the yield limit ($Bn \to \infty$), the flow is highly localised to some channels while other parts of the fluid are quiescent. At this limit, $\Delta P / L \sim Bn^n$ where $n$ is equal to one for ``simple" viscoplastic fluids \cite{chaparian2019porous}. However, as depicted in Fig.~\ref{fig:pressuredrop}, for elastoviscoplastic fluids, $n$ is less than one which renders the critical pressure gradient (or the inverse of the critical yield number; see \cite{chaparian2021sliding,chaparian2024percolation}) to be smaller in elastoviscoplastic fluids compared to the viscoplastic fluids.

As previously observed in numerous number of studies on viscoelastic liquids, high elasticity of the fluid (i.e.~high $Wi$) results in unsteadiness, chaos and finally elastic turbulence in the inertialess regime \cite{Steinberg2021}; see \cite{de2017viscoelastic} in the context of porous media flows. In elastoviscoplastic fluid flows in a periodic cell (i.e.~model porous media), we previously observed a transition to time-dependent oscillations at $Bn=10$ and $Wi\approx0.5$ \cite{de2018elastoviscoplastic}, using a volume-penalization IBM approach with $>700$ grid points over each cylinder surface. In elastoviscoplastic fluids, the elastic behaviour is characterised by $Wi \times Bn$ as evidenced both computationally and experimentally in our previous studies \cite{chaparian2019adaptive,Izbassarov2020,Izbassarov2021,villalba2023}. Hence, even at relatively small Weissenberg numbers, if the Bingham number is sufficiently high, the unsteadiness appears; here we quantified this behaviour in Figs.~\ref{fig:Pre_Ins} and \ref{fig:Pre_Ins_NotC} where the pressure drop is shown as a function of time. Also, the effect of porosity has been discussed: at low solid volume fractions (e.g.~$\phi=10\%$), the flow is steady at low $Bn$ and $Wi$, while increasing the volume fraction to $\phi=20\%$ or $30\%$ makes the flow unsteady at the same low $Wi$ and $Bn$. In summary, the pressure drop is smaller and time-independent (steady) when the medium is more porous and the flow is at low $Bn$ and $Wi$ numbers. Conversely, for larger values of the Bingham number, Weissenberg number and the solid volume fraction, the pressure drop becomes considerably time-dependent. This is in line with studies of viscoelastic fluids in porous media, see e.g. \cite{Kumar2022}. Recent studies \cite{Haward2021} have suggested that the number of stagnation points in the flow affects the onset of elastic turbulence. The number of stagnation points naturally increases with the solid volume fraction, but also depends on the precise geometric configuration in the disordered porous medium \cite{Walkama2020,Haward2021}. Elastic turbulence results in a sudden increase in drag above a certain Weissenberg number threshold in viscoelastic flows through porous media \cite{Browne2021}. Hypothetically, if this were to happen in elastoviscoplastic flows as well when $Wi \times Bn$ becomes very large, then our conclusion of drag reduction by elasticity may change. Characterising the elastoviscoplastic turbulence in porous media is, however, beyond the scope of the present work.

Due to the high cost of numerical simulations of elastoviscoplastic fluid flows, here we focused on unravelling general features of this kind of fluids in porous media flows. Future studies are required to delve into the details and also the bulk transport properties which hopefully can yield general Darcy law for such kind of fluids. Thus, we only scratched the surface here which can be a baseline for future studies to extend our work in various directions, for example, focusing on more realistic topologies (e.g.~digital rocks), more sophisticated rheological models (e.g.~thixo-elasto-visco-plastic behaviours), or inertial regimes. Also, in industrial applications, the porosity of the medium can be very small such as ``digital rocks" but for 2D randomised mono-dispersed obstacles, we cannot reach that limit. This remains for future studies to be addressed.


\bmhead{Acknowledgments} The authors acknowledge computing time provided by SNIC (Swedish National Infrastructure for Computing) and the National Academic Infrastructure for Supercomputing in Sweden (NAISS).

\bmhead{Funding} This work received funding from the European Research Council (Grant no.~ERC-2019-StG-852529, MUCUS), and by the Swedish Research Council grant No.~VR2021-04820.

\bmhead{Consent for publication} All authors gave their final approval for publication.

\bmhead{Authors' contributions} S.P. and E.C. conducted the numerical simulations and analysis. All the authors contributed in writing the manuscript and editing.

\bmhead{Availability of data and materials} The datasets can be accessed via request. Contact the corresponding author.

\vspace{5pt}

{\noindent \bf Declarations}

\bmhead{Conflict of interest} The authors report no conflict of interest.

\bmhead{Competing interests} The authors report no competing interests.

\bmhead{Ethical approval consent to participate} Not applicable.

{\color{red}
\section*{Appendix A: A different possible scaling}

The non-dimensional numbers which are introduced in \S \ref{section:Problem_setup} below Eq.~\ref{eq:ucd}, can be redefined in the way that discussed in \cite{thompson2016viscoplastic}. Indeed, rather than using the characteristic viscous stress, characteristic total stress ($\hat{\tau}_c = \hat{\tau}_y + \hat{\mu} \hat{U} / \hat{R}$) can be used to make the yield stress and the pressure drop non-dimensional, where $\hat{\mu}$ is the plastic viscosity in the case of Bingham fluid and is the total viscosity in the case of elastoviscoplastic fluid: 

\[
Bn^* = \frac{\hat{\tau}_y}{ \hat{\tau}_y + \hat{\mu} \hat{U} / \hat{R} } = \left( 1 + \frac{1}{B} \right)^{-1},
\]

\[
\left( \Delta P / L \right)^* = \frac{\Delta \hat{P} / \hat{L}}{ \hat{\tau}_y / \hat{R} + \hat{\mu} \hat{U} / \hat{R}^2 } = \frac{\Delta P/L}{1+B}.
\]

Please note that, $Bn^*$ is called plastic number in the previous works.

\begin{figure}[h]
\centering
\includegraphics[width=\linewidth]{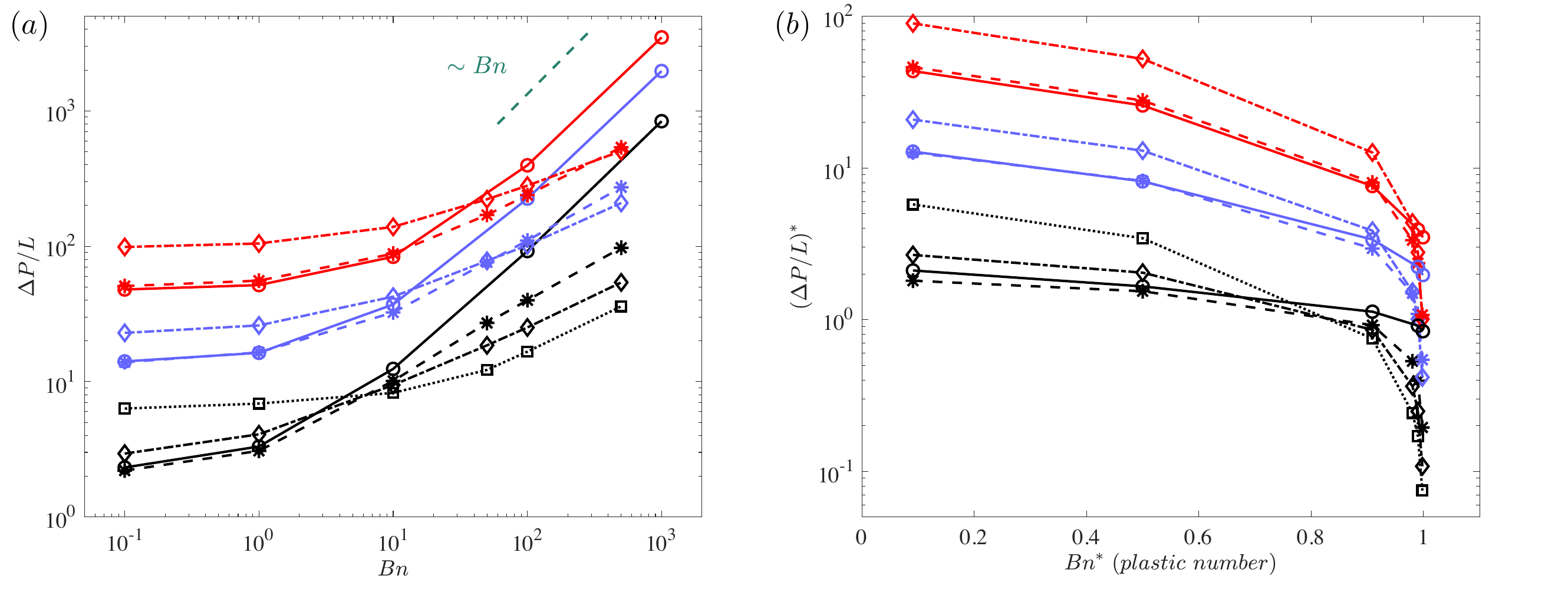}
\captionsetup{justification=centering}
\caption{\label{fig:Roney} Non-dimensional pressure drop as a function of (a) Bingham number; (b) plastic number. The colour and symbol interpretations are the same as Fig.~\ref{fig:pressuredrop}.}
\end{figure}

}

\bibliography{Revision2.bib}

\end{document}